\documentclass[journal]{IEEEtran}

\usepackage{latexsym}%
\usepackage{cite}%
\usepackage{amsmath}%
\usepackage{amssymb}%

\usepackage{bm}

\usepackage{dcolumn}

\usepackage{ifpdf}
\ifpdf 
  \usepackage[pdftex]{graphicx}
  \DeclareGraphicsExtensions{.pdf,.png,.jpg,.jpeg,.mps}
  \usepackage{pgf}
\else 
  \usepackage{graphicx}
  \DeclareGraphicsExtensions{.eps,.bmp}
\fi
\usepackage{epic,bez123}

\begin{document}

\title{ Error Correction for NOR Memory Devices with Exponentially Distributed Read Noise }

\author{{Daniel~L.~Miller}
\\Micron Israel, P.O. Box 1000, Qiryat-Gat 82109, Israel (email: dmillera@micron.com)
}


\maketitle

\begin{abstract}
\boldmath
   The scaling of high density NOR Flash memory devices with 
   multi level cell (MLC)  hits the reliability break wall
   because of relatively high intrinsic bit error rate (IBER).
   The chip maker companies offer two solutions to meet 
   the output bit error rate (OBER) specification: 
   either partial coverage with error correction code (ECC)
   or data storage in single level cell (SLC) 
   with significant increase of the die cost.
   The NOR flash memory allows to write information in small portions, 
   therefore the full error protection becomes costly 
   due to high required redundancy, e.g. $\sim$50\%.
   This is very different from the NAND flash memory
   writing at once large chunks of information; 
   NAND ECC requires just $\sim$10\% redundancy.
   This paper gives an analysis of a novel error protection 
   scheme applicable to NOR storage of one byte. 
   The method does not require any redundant cells, but assumes 5th program level.
   The information is mapped to states in the 4-dimensional 
   space separated by the minimal Manhattan distance equal 2.
   This code preserves the information capacity: one byte occupies four memory cells.
   We demonstrate the OBER $\sim$ IBER$^{3/2}$ scaling law, 
   where IBER is calculated for the 4-level MLC memory.
    As an example, the 4-level MLC with IBER $\sim10^{-9}$, 
    which is unacceptable for high density products,
    can be converted to OBER $\sim10^{-12}$.
   We assume that the IBER is determined by the exponentially distributed read noise.
   This is the case for NOR Flash memory devices, since the exponential tails are typical for 
   the random telegraph signal (RTS) noise and for most 
   of the charge loss, charge gain, and charge sharing data losses. 
\unboldmath
\end{abstract}


\section{ Introduction }

\IEEEPARstart{T}{he} Flash memory devices store information in array of 
memory cells, with every cell (memory transistor) programmed to a certain 
level (value) of the threshold voltage $V_t$. The modern technology offers 
two types of the Flash memory devices: NOR and NAND. The NAND memory operates 
by large chunks of data stored with density $\sim10^{11}$ bit/cm$^2$ 
and relatively high IBER $\sim10^{-2}$. The NOR memory allows WRITE/READ 
of a single bit/byte; the data is stored with density $\sim10^{10}$ 
bit/cm$^2$ and relatively low IBER $\sim10^{-9}$. The state of the art error 
correction codes (ECC) were developed for NAND memories, see e.g.  \cite{Gregori2003,eccbook,BCH-Hamming,Reed-Solomon}.
These codes are capable to repair multiple errors and reduce the OBER to $\sim10^{-14}$. 
The modern ECCs are not applicable to the NOR Flash memory devices, 
because the efficiency of the ECC increases with amount of the data 
written at once  \cite{eccbook,patterns} which is above Kbyte for NAND. 
However, amount of the data written at once is single bit/byte for NOR. 
For example, the Hamming code correction of 
a single error in 4 cells storing 1 byte (2 bits/cell) requires 2 more redundant cells 
and 50\% die size increase. Stringent requirements to OBER 
limit the scaling of the NOR Flash memory devices: die size gain due to scaling 
of the cell size is wasted for accommodation of redundant cells.

Angelo Visconti patented the idea of adding redundant program levels 
to the NOR Flash memory cell along with applying an error correction code
 \cite{angelo-2003}. In particular, he proposed to map the data to alphabet 
of size 5, write to 5 levels per cell, and protect the information by 
the relevant Hamming code. For example, 64 bits of the data are written into 28 cells, 
then 4 redundant cells allow correcting an error in any of 32 cells.
This method preserves the density of 2 bits/cell, 
however it requires the WRITE operation of at least 8 bytes at once.

The READ error in Flash memory device occurs because of the overlap 
of $V_t$ distributions between neighbor program levels. The IBER $\sim10^{-10}$
in NOR Flash memory devices means that the overlap is weak, and the IBER 
is determined by the tail of the $V_t$ distribution. The effect of trapping and 
de-trapping of charges is responsible for an exponential shape of these tails  \cite{Traps-Seminal}.
The exponential shape of the $V_t$ distribution is typical 
for RTS noise, cycling effects on charge retention\cite{Shape-Onur}, 
and cell interference  \cite{ECC-RTN}. The slope of the exponential distribution 
depends on many factors including channel doping  \cite{RTS-Ghetti}, 
the memory usage model, etc. 
This is quite opposite to the NAND, where IBER is high and the Gaussian $V_t$ distribution is adequate.

This paper considers an alternative approach to error correction in the NOR Flash memory devices.
The idea is to add program levels similarly to  \cite{angelo-2003}, however to encode the information 
by maximizing the minimum Manhattan distance  \cite{SL}. The Manhattan metrics is optimal 
for systems with exponential noise, because the error probability becomes the exponential function of the 
Manhattan distance between neighbor states.
The method is closely related to non-binary coding in the Lee metrics  \cite{lee-1958,ulrich-1958,Golomb-Welch-1970}. 
Recent developments in polyomino (cross-polytopes) tiling make the idea 
attractive for Flash memory design  \cite{Schwartz-tiling,etzion-yaakobi,Horak-AlBdaiwi}.

Below we present a calculation of the gain of storage reliability of 
eight bits in a system having 4 memory cells. As the reference we take
the system with 4 program levels (two bits per cell, $B_0=2$). 
The additional 5th program level will increase the information 
capacity of the system to $\log_2(5)=2.3$ bits per cell. 
Let the word $x_1\ldots x_4$, where $x_j\in \{0,1,2,3,4\}$,
describe the state of the Flash memory system; the $j$-th cell is programmed to 
$x_j$-th program level. The coding of the
information with minimum Manhattan distance equal two  \cite{SL} 
\begin{equation}
   x_1 + x_2 + x_3 + x_4 = 0 \mod 2
\label{eq:md2}
\end{equation}
will reduce the information capacity to 
\begin{equation}
   B = (1/4)\log_2({5^4+1\over2})=2.07 > B_0=2\;.
\end{equation}
Therefore, the modified system with 5 levels per cell and non-binary coding 
will have enough information capacity to store 8 bits in 4 cells.

The read of information from the $j$-th Flash memory cell is done by
comparison of the device threshold voltage with the reference value stored 
in the reference memory cell and determination of $\tilde x_j$. 
With high probability the parity is satisfied $\tilde x_1 + \tilde x_2 + \tilde x_3 + \tilde x_4 = 0 \mod 2$, 
and the READ is correct;
otherwise there is an error and soft sensing \cite{SoftSensing} for error correction is required. 
The periphery circuit measures the threshold voltages of all 4 cells and 
searches for nearest (in Manhattan distance) word 
satisfying (\ref{eq:md2}). The probability of wrong error correction is 
of the same order as the probability of two errors, see calculations in Sec. \ref{sec:5levels}. 
Therefore, the logarithm of the inverse OBER can be increased as much as 50\% by adding the 5th program 
level to the 4-level memory cell. As an example, the 4-level MLC with IBER $\sim10^{-9}$, 
which is unacceptable for high density products,
can be converted to OBER $\sim10^{-12}$, see Table \ref{table:1}.

\input{NOR5_4Levels.TpX}

\input{NOR5_5Levels.TpX}

\section{ The bit error rate of an unprotected system }

Observe $N$ memory cells in a given memory device.  Then program these $N$ memory cells to 
predefined threshold voltage levels $\{L_0,\ldots,L_3\}$ see Fig. \ref{fig:4levels}(a).
The program state $A'$ of memory system is
\begin{equation}
               A'=\{V'_1,\ldots,V'_N\}\;,
\end{equation}
where the threshold voltage $V'_j$ of the $j$-th memory cell is uniformly distributed around $x_j$-th program level $L_{x_j}$
\begin{equation}
   f_p(V) = \left\{
   \begin{array}{lr}
              1/W, & L_x-W/2 < V < L_x+W/2 \\
              0 , & \text{otherwise}
   \end{array}\right.
\end{equation}
The width $W$ of the program distribution is typically a function of the program speed. Faster programming leads to 
the wider program distribution and larger $W$. 

The READ operation of the Flash memory device cannot reproduce exactly the state $A$. The state is distorted by the read
noise ( typically the RTS of the read current ) and by the data retention issues, see Fig. \ref{fig:4levels}(b).
The periphery circuit reads the state 
\begin{equation}
               A=\{V_1,\ldots,V_N\}\;
\end{equation}
of the memory system. The distribution of the threshold voltage $V_j$ acquires the exponential tail
\begin{equation}
   f_R(V) = 
\left\{
   \begin{array}{l}
              (1-T)/W, L_x-W/2 < V < L_x+W/2 \\
              a T e^{-2a (V - L_x- W/2) }, V > L_x+W/2 \\
              a T e^{-2a (L_x- W/2 - V) }, V < L_x-W/2 \\
   \end{array}\right.
\end{equation}
where $T$ is the fraction of the cells in the tail, and $1/2a$ is the slope of the distribution.

The READ operation is simply comparison of the threshold voltage of the memory cell with the 
the threshold voltages of the reference cells. The threshold voltages of the reference cells
is positioned in the middle of the level-to-level spacing.
The READ operation finds values of the word $\{x_j\}$ from $\{V_j\}$
\begin{equation}
    x_j = \text{Round} { V_j-L_0 \over \Delta_0 + W }\;,
\label{eq:margin_sense}
\end{equation}
where rounding is performed to the nearest integer.

The probability of the read error of one cell is 
\begin{eqnarray}
    P_0 &=& P(V > L_{x}+\Delta_0/2+W/2)  
\nonumber\\
    &+& P(V < L_x - W/2 -\Delta_0/2) 
\nonumber\\
    &=& 2\int_{\Delta_0/2}^\infty aT e^{-2aV} dV = T e^{-a\Delta_0}\;.
\end{eqnarray}
The probability of error in reading of $N$ cells per bit of information (IBER) is therefore 
\begin{equation}
   E_0 = { 1 - ( 1-P_0 )^N \over NB_0 } \approx {1\over 2} T e^{-a\Delta_0}\;.
\label{eq:bareBER}
\end{equation}
The key parameter here is $a\Delta_0$, which is the level spacing times the exponent of noise distribution.
For systems with no error correction and large volumes of the data storage, $E_0\sim10^{-12}$ is required.

\section{ The bit error rate of five level system with non-binary coding }
\label{sec:5levels}

Let's program $N=4$ memory cells to 5
predefined threshold voltage levels $\{L_0,\ldots,L_4\}$ see Fig. \ref{fig:5levels}(a).
The program state $A'$ of memory system is
\begin{equation}
               A'=\{V'_1,\ldots,V'_4\}\;,
\end{equation}
where the threshold voltage $V'_j$ of the $j$-th memory cell is uniformly distributed around $x_j$-th program level $L_{x_j}$,
and the word $\{x_j\}$ satisfies (\ref{eq:md2}).

The READ operation becomes different from the margin sensing (\ref{eq:margin_sense}). The READ will require
soft sensing for the error correction:
\begin{enumerate}
\item
Assume that the system read the state $A=\{V_1,\ldots,V_4\}$.
\item
Read the encoded word $\{\tilde x_j\}$ by margin sense (\ref{eq:margin_sense}).
\item
If the parity condition (\ref{eq:md2}) is satisfied for $\tilde x_j$, then the READ operation was correct.
The word $\{\tilde x_j\}$ is the READ of the encoded word $\{x_j\}$.
\item
If the parity condition (\ref{eq:md2}) is not satisfied for $\tilde x_j$, then the error correction is required.
\item
Take all allowed states $B=\{L_{y_1},\ldots,L_{y_4}\}$, such that $\sum y_j=\sum \tilde x_j\pm1$, and define the distance 
\begin{equation}
    |AB|=\sum_j |V_j-L_{y_j}|\;.
\label{eq:L1_metrics}
\end{equation}
\item
Find the state $B$, such that $|AB|$ is minimum.
\item
The word $\{y_j\}$ is the READ of the encoded word $\{x_j\}$.
\end{enumerate}
This is essentially ``soft-decision sensing'', as described in  \cite{SoftSensing}.

Three types of errors occur in this model. If the threshold voltages of two cells are found 
more than $(\Delta+W)/2$ away from $L_x$, then the parity check will pass, and the error will not be detected.
The probability of this error and the corresponding OBER are
\begin{eqnarray}
    P_2^{i} &=& T^2 e^{-2a\Delta}\;,
\\
    E_2^{i} &=& {3\over 8}  T^2 e^{-2a\Delta}\;.
\end{eqnarray}
If the threshold voltage of $j$-th cell moves more than $(2\Delta+W)/2$ away from $L_{x_j}$, then the error correction
algorithm will converge to $y_j=x_j\pm2$, see Fig. \ref{fig:5levels}(b). 
The probability of this error and the corresponding OBER are
\begin{eqnarray}
    P_2^{ii} &=& T e^{-2a\Delta-aW}\;,
\\
    E_2^{ii} &=& {1\over 2}  T e^{-2a\Delta-aW}\;.
\end{eqnarray}
The third type of error occurs when 
\begin{equation}
      \exists j\ne k \quad y_j=x_j+1\;,\quad y_k=x_k-1\;.
\end{equation}
In terms of the read threshold voltages the condition for the error is 
\begin{equation}
    |V_j-L_{x_j+1}|+|V_k-L_{x_k-1}| < |V_j-L_{x_j}|+|V_k-L_{x_k}|\;,
\label{eq:error_iii}
\end{equation}
which is derived from (\ref{eq:L1_metrics}).
In the relevant range 
\[
     L_{x_j} <  V_j < L_{x_j+1}\;,\quad L_{x_k-1} <  V_k < L_{x_k}
\]
it becomes simplified to 
\begin{equation}
    L_{x_j+1}-V_j <  L_{x_k} - V_k \;.
\end{equation}
The probability of this event is better expressed in terms of the deviation variables 
$\varepsilon_j=|V_j-L_{x_j}-W/2|$ and $\varepsilon_k=|V_k-L_{x_k}-W/2|$,
\begin{eqnarray}
   P_2^{iii} &= & 
         a^2T^2 \int_{0}^{\Delta} d\varepsilon_j 
       \int^{\Delta}_{\Delta - \varepsilon_j} d\varepsilon_k 
        e^{-2a\varepsilon_j-2a\varepsilon_k}
\\
      &\approx&  {1\over 2} a\Delta T^2 e^{-2a\Delta}\;.
\end{eqnarray}
The sum over pairs of cells and the normalization per number of stored bits gives
\begin{equation}
   E_2^{iii}= {1\over NB_0} \sum_{k\ne j}P_2^{iii} = {3\over 4} a\Delta T^2 e^{-2a\Delta}\;.
\label{eq:errormisplace}
\end{equation} 
The total BER of five level design with non-binary ECC becomes
\begin{eqnarray}
   E_2 &=& E_2^{i} + E_2^{ii} + E_2^{iii}
\nonumber\\
   &=&  {3\over 4} T^2 \left(a\Delta+{1\over 2}\right) e^{-2a\Delta} + {1\over 2}  T e^{-2a\Delta-aW}
\end{eqnarray}
This must be compared with the probability $E_0$ of the error in the 4-level system given by (\ref{eq:bareBER}).
Assuming that the 5th level was added without pushing out $L_0$ and $L_3$, as in Figs. \ref{fig:4levels},\ref{fig:5levels},
we get the condition
\begin{equation}
    4 \Delta  = 3 \Delta_0 - W \;.
\label{eq:4to5}
\end{equation}
The generic scaling law for the OBER is (assume $W\ll\Delta_0$)
\begin{equation}
   E_2 \sim e^{-2a\Delta} \sim e^{-(3/2)a\Delta_0} \sim E_0^{3/2}\;.
\label{eq:scaling}
\end{equation}
In practical calculation the factors $T$ and $e^{-aW}$ are essential, see the next section.

\begin{table}
\caption{Bit error rate gain by extra level in $B=2$ bit/cell memory system. 
         The write is restricted to 8 bits at a time (4 cells). }
\label{table:1}
\centering
\setlength{\tabcolsep}{12pt}
\begin{tabular}{c|c|c|c|c}
\hline
$e^{-a\Delta_0}$
      &$T$   &$E_0$ &$E_2$&$E_2/E_0$\\
\hline
1.E-02&1.E-03&5.E-06&7.E-08&1.E-02\\
1.E-02&1.E-05&5.E-08&5.E-10&1.E-02\\
1.E-02&1.E-07&5.E-10&5.E-12&1.E-02\\
1.E-02&1.E-09&5.E-12&5.E-14&1.E-02\\
1.E-03&1.E-03&5.E-07&3.E-09&7.E-03\\
1.E-03&1.E-05&5.E-09&5.E-12&1.E-03\\
1.E-03&1.E-07&5.E-11&5.E-14&1.E-03\\
1.E-03&1.E-09&5.E-13&5.E-16&1.E-03\\
1.E-04&1.E-03&5.E-08&4.E-10&7.E-03\\
1.E-04&1.E-05&5.E-10&9.E-14&2.E-04\\
1.E-04&1.E-07&5.E-12&5.E-16&1.E-04\\
1.E-04&1.E-09&5.E-14&5.E-18&1.E-04\\
\hline
\end{tabular}
\end{table}

\section{Practical examples.}

\label{sec:M5y2}

It is quite common in high reliability NOR Flash memory devices to have very 
low fraction of cells suffering from the data retention issues, $T\sim $1E-6, $T\ll e^{-aW}$.
In this case
\begin{equation}
   E_2 \approx E_2^{ii} = {1\over 2}  T e^{-(3/2)a\Delta_0-(1/2)aW} = e^{-a\Delta_0/2-aW/2} E_0
\label{eq:e2ii}
\end{equation}
The other possibility is to have relatively narrow program distributions for better read window, $e^{-aW}\ll T$:
\begin{equation}
   E_2 \approx E_2^{iii} =  
   {3\over 2} { 3 a\Delta_0 - aW \over 4} T  e^{-a(\Delta_0-W)/2} E_0\;.
\label{eq:e2iii}
\end{equation}
In (\ref{eq:e2ii}), (\ref{eq:e2iii}) the spacing $\Delta$ of 5-level system is expressed in terms of $\Delta_0$ 
and $W$ by making use of (\ref{eq:4to5}).

The proposed method yields a substantial gain in OBER performance for practical applications,
see Table \ref{table:1}. For estimation purpose we put $W\sim \Delta_0$
and the OBER/IBER ratio becomes
\begin{equation}
   E_2/E_0 \approx e^{-a\Delta_0} + {3\over 4} a\Delta_0 T \;,
\end{equation}
and use this formula for typical values of $T$ and $e^{-a\Delta_0}$. The 5-level non-binary coding allows to reduce 
the OBER by 2-4 orders of magnitude. This method allows the technology progress from gigabit NOR parts to tenth gigabit
products. The one-byte non-binary ECC discussed in this paper 
can be combined with global ECC coverage in cases where the data is streamed in many bytes.

For the systems with purely exponetial read noise, $W=0$, $T=1$, the scaling law (\ref{eq:scaling}) becomes accurate
and can be used for OBER calculation. In this case the $E_0\sim10^{-10}$ is reduced to $E_2\sim10^{-15}$.

For the applications with the relatively high IBER, the 5-level method does not gain enough OBER.
The reliability improvement can be achieved by using more complicated parity rules  \cite{SL} with 
higher Manhattan distance \cite{lee-1958} and more levels per cell. For example, the minimum Manhattan distance equal 3 can be
 achieved by packing the cross-polytopes of the radius 1. For the system of 4 memory cells, the volume of the 4-dimensional
 cross-polytope is 9. Therefore, the 7 levels per cell are required to preserve the information capacity of the system
$(1/4)\log_2(7^4/9)=2.014>2$.

\begin{IEEEbiography}[{\includegraphics[width=1in,height=1.25in,clip,keepaspectratio]{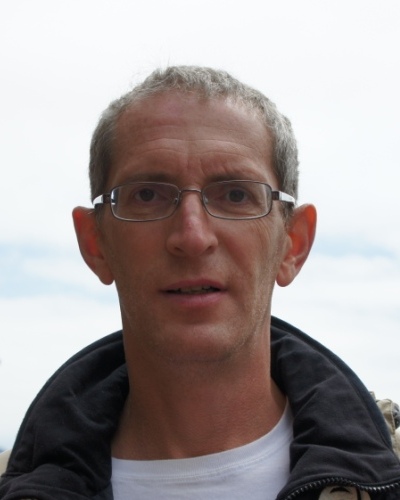}}]{Daniel L. Miller}
holds a Ph.D. in theoretical physics from the Hebrew univ. in Jerusalem. 
He worked in the Weizmann Inst. of Science 1996-1999 in the field of quantum chaos. 
In 1999 he joined Intel and worked 
for 10 years as a device engineer on sustaining and developing 
numerous technologies. 
He has been with Micron Israel since 2010, working in R\&D on 45nm NOR devices.
\end{IEEEbiography}

\end{document}